\documentstyle[aps,12pt]{revtex}

\input psfig.tex

\begin{document}
\title{The Anomalous Chromomagnetic Dipole Moment of Top Quark}
\author{R. Mart\'{\i}nez$^1$, J.-Alexis Rodr\'{\i}guez$^{1,2}$ and
M. Vargas$^1$}
\address{$^1$Depto. de F\'{\i}sica,
Universidad Nacional, Bogot\'{a}, Colombia\\
$^2$Centro Internacional de F\'{\i}sica, Bogot\'a, Colombia}
\date{}
\maketitle

%\vspace{3cm}

\begin{abstract}
We calculate the one loop corrections to the chromomagnetic
dipole moment of the top
quark 
in the framework of the Standard Model, the two Higgs doublet model
and the Minimal Supersymmetric Standard Model. In the Standard Model we
consider the QCD and the electroweak corrections generated by
gluon and Higgs boson exchanges, respectively. In the Minimal
Supersymmetric Standard Model, we calculate the SUSY- QCD
corrections, including the mixing effects between the stop
coming from the left- and right-handed sectors. We obtain that the
Standard Model contribution is of the order of $-0.004 \leq \delta
\kappa_g^t \leq -0.001$, and the values for diferents scenarios are of the
order of $-0.001 \leq \delta \kappa_g^t \leq -0.01$.
\end{abstract}
%\newpage

Up to now, the Standard Model (SM) gives an adequate description of the
data that involve the strong and electroweak interactions \cite{salam} .
In particular, the discovery
of the top quark at Fermilab by CDF and D0 \cite{cdf}, with a mass of
$174\pm 6$
GeV within the range allowed by the
high precision tests of the
SM \cite{bernabeu}, has showed that the SM is a good scheme that describes
physics at these
energies. However, the SM is not considered the final theory, and
physicists
think in the existence of new physics. In particular,
since the fermion mass generation may be closely related to the
electroweak
symmetry breaking, one expects that some residual effects of the breaking
mechanism to appear
in accordance with the generated mass \cite{peccei}.

The top quark as the heaviest fermion in the model with a mass of the
order of
the  symmetry breaking scale, opens a window to new physics, and one
natural
scenario
is the modification of its interactions with the known gauge bosons.
Future experimental runs will increase the number of produced $t\overline{t\text{ }}
$ pairs, allowing for the comparison of measured differential cross
sections with theoretical predicitons of the many different models. In
particular, one
hopes
to obtain detailed information on the couplings of the top quark 
\cite{larios}.

Some $ SU(2)_L \otimes U(1)_Y$ invariant operators, which represent new
physics efects, give rise to 
couplings of the dipole moment type \cite{larios} \cite{wyler}. In the
case of
strong interactions, there are
some CP-invariant operators that contribute to the anomalous
chromomagnetic dipole moment of
the top quark, $\delta\kappa_g^t$.
Although there are stringent experimental bounds on anomalous contributions
to dipole moments of light fermions, it is not unreasonable to expect large
anomalous moments of the heavy top quark \cite{nacht}. 

The chromomagnetic
dipole moment deserves  special interest because this coupling is 
crucial in the
production of top quark pairs by the anihilation processes $e^{+}e^{-}\to
\bar{t}t$
and $\bar{q}q\to \bar{t}t$ \cite{silver}. For instance, the top quark
anomalous couplings
can modify the energy distribution of the extra gluon jet associated with top
pair productions, and the production of top quark pairs in association
with
a hard gluon in future $e^{+}e^{-}$ experiments provides a good
opportunity to
probe the anomalous $ttg$ couplings \cite{rizo}.

This possibility has
been considered, exploring the implications of scenarios like the SM \cite
{atwood}, the two-Higss-doublet model (2HDM) \cite{stange} and the
different
contributions of the Minimal Supersymmetric Standard Model (MSSM)
\cite{li}.
More recently,
indirect bounds for the anomalous vertex have been obtained from
known
experiments, such as CLEO for B decays, which was used to
bound the magnetic and chromomagnetic dipole moment of the top quark within the context
of a decoupled effective Lagrangian \cite{martinez}.

To naively estimate the contribution of new physics effects to
$\delta\kappa^t_g$, we can
remove the gluon line from the vertex diagram, and then get the mass
term for the
top quark. By dimensional
analysis this can be written as 
\begin{equation}
\delta\kappa_g^t\approx\frac{m_t^2}{\Lambda^2} \; \; ,
\end{equation}
where $\Lambda$ is the scale of new physics. If new physics appears at
$\Lambda=1$ TeV, then $\delta\kappa_g^t\approx 0.031$. In the literature
some models that give non-zero $\delta\kappa_g^t$ 
value have been considered. Technicolor models \cite{tecni} predict
\begin{equation}
\delta\kappa_g^t\approx\frac{2 m_t^2}{m_{\omega}^2}\; \; ,
\end{equation}
where $m_{\omega}$ is the mass of the techniscalar exchanged in the loop.
For $m_{\omega}\geq 0.5$ TeV, $\delta\kappa_g^t\leq 0.25$ \cite{atwood}.

Our goal in this paper is to calculate
explicitly the anomalous chromomagnetic dipole moment of
the top quark in the framework of the SM, the 2HDM,
and  supersymmetric QCD.

The first point that  we wish to make
in this paper is the status of the SM with respect to the $%
ttg $ anomalous chromomagnetic dipole moment which is defined as 
\begin{equation}
{\cal L}=i \delta\kappa_g^t\frac{g_s}{2 m_t} \;\bar{t}\; T^a\;
\sigma_{\mu\nu}
q^{\nu}\; t\; G^{\mu, a}
\end{equation}
where $g_s$ and $T_a$ are the usual $SU(3)_C$ coupling and operator.
In the case of QCD, the
anomalous factor comes from the calculation at one loop level of the vertex $%
ttg$ by considering the exchange of gluons, and the result is 
\begin{equation}
\delta \kappa _g^t=-{\frac{\alpha _s (m_t)}{12\pi}}  \; \; .
\end{equation}
We found that the top quark contribution into the loop is
decoupling\cite{carazone},
and the dependence of $\delta \kappa _g^t$ with respect to $m_t$ is
determined by the running coupling constant.

On the other hand, the most
important SM contribution to the anomalous chromomagnetic dipole coupling
is due to the
exchange of neutral Higgs boson and would-be Goldstone boson. The
contribution
of this sector can be expressed as:
\begin{equation}
\delta \kappa _g^t={\frac{-\sqrt{2}G_Fm_t^2}{8\pi ^2}}\left[ \int_0^1dx{%
\frac{x-x^3}{x^2-a(m_h^2)x+1}}+\int_0^1dx{\frac{-x^3+2x^2-x}{x^2-a(m_Z^2)x+1}%
}\right]
\end{equation}
with $a(m_i^2)=2-m_i^2/m_t^2$, $m_h$ the Higgs mass and $m_Z$ the  $Z$ mass.
The above contribution is showed as a function of the
Higgs mass in figure 1 (solid line). Also we display the sum
of QCD and SM contributions (dashed line). The
horizontal dashed
line is the QCD contribution from eq.(4).

Our next task is to calculate the contribution of the 2HDM
\cite{hunter} to the anomalous chromomagnetic dipole moment of the top
quark.
This model preserves
the $SU(2)_L \otimes U(1)_Y$ SM symmetry,
and includes two scalar isodoublets $\Phi _i$.
The Yukawa couplings of the scalars and fermions are constrained by the
requirement that there are no flavor changing neutral interactions at
tree level \cite{paschos}. There are
two possible arrangements for the Yukawa couplings: either one may couple all
of the fermions to a single Higgs doublet (model I), or one may couple $%
\Phi _1$ to the right-handed down-type quarks, and $\Phi _2$ to the
right-handed up-type quarks. This is the so-called model II, which
includes the scalar sector of the MSSM as a particular case. Here, we will
consider the
latter
possibility. The Yukawa Lagrangian for the quark sector has the form 
\begin{equation}
-L_Y=\sum_{i,j=1}^3 {\cal D}_{ij}\overline{q}_{i,L}\Phi
_1d_{j,R}+\sum_{i,j=1}^3 {\cal U}_{ij}%
\overline{q}_{i,L}\tilde{\Phi}_2u_{j,R}
\end{equation}
where $\overline{q}_{i,L}$ are the left-handed quark doublets, ${\cal U}$
and ${\cal D}$ are the up and down quark mass
matrices, respectively.

In this case, the
vacuum expectation values of the Higgs fields $\Phi_i$ can be chosen
to be both real and positive: 
\begin{equation}
\left\langle \Phi _1\right\rangle =%
%TCIMACRO{\binom 0{\nu _1/\sqrt{2}}}
%BeginExpansion
{0 \choose \nu _1/\sqrt{2}}
%EndExpansion
\; \; \;
\text{   ,   }\left\langle \Phi _2\right\rangle =%
%TCIMACRO{\binom 0{\nu _2/\sqrt{2}}}
%BeginExpansion
{0 \choose \nu _2/\sqrt{2}}
%EndExpansion
\;\;
\text{.}
\end{equation}
By diagonalizing the Higgs mass matrices one finds the physical Higgs
states. In the charged sector, there are a charged would-be Goldstone boson and a
charged scalar Higgs, 
\begin{eqnarray}
G^{\pm } &=&\phi _1^{\pm }\cos \beta +\phi _2^{\pm }\sin \beta \;\; \text{   ,} \\
H^{\pm } &=&\phi _2^{\pm }\cos \beta -\phi _1^{\pm }\sin \beta    
\nonumber
\end{eqnarray}
where the mixing angle $\beta $ is given by $\tan \beta =\nu _2/\nu _1$.
Similarly, the imaginary parts of the neutral components mix to form a
neutral would-be Goldstone boson and a CP-odd  scalar Higgs, 
\begin{eqnarray}
G^0 &=&\Im \phi _1^0\cos \beta +\Im \phi _2^0\sin \beta \;\; \text{,}
\\
A^0 &=&\Im\phi _2^0\cos \beta -\Im\phi _1^0\sin \beta \;\; \text{.}
\nonumber
\end{eqnarray}
Finally, the real parts of the components mix to form a pair of
neutral CP-even scalars Higgs, 
\begin{eqnarray}
H^0 &=&\sqrt{2}((\Re \phi _1^0-\nu _1)\cos \alpha +(\Re \phi
_2^0-\nu _2)\sin \alpha )\;\; \text{,} \\
h^0 &=&\sqrt{2}((\Re \phi _2^0-\nu _2)\cos \alpha -(\Re \phi
_1^0-\nu _1)\sin \alpha )  \nonumber
\end{eqnarray}
where the mixing angle $\alpha$ depends on the Higgs potential parameters.

The quantity $\tan \beta $ plays a central role in the theory because
the Yukawa couplings are often proportional to either $\tan \beta $ or $\cot
\beta $, 
\[
\begin{tabular}{p{1.in}p{2.in}p{1.in}}
Higgs & $\overline{u}u$ & $\overline{d}d$ \\ 
$H^0$ & $m_u\sin \alpha /\sin \beta $ & $m_d\cos \alpha /\cos \beta $ \\ 
$h^0$ & $m_u\cos \alpha /\sin\beta $ & $-m_d\sin \alpha /\cos \beta $ \\ 
$A^0$ & $m_u\cot \beta \gamma _5$ & $m_d\tan \beta \gamma _5$%
\end{tabular}
\]
In the above, we summarize the Yukawa couplings in  model II
(we have omitted the overall factors $-ig/2m_W$ for $H^0,h^0$ and
$-g/2m_W$
for $A^0$).
We may note that $0\leq \beta \leq \pi/2$  since we take $\nu
_1,\nu _2>0.$ In Appendix A of ref. \cite{gunion} it is shown that this
implies
$-\pi /2\leq \alpha \leq 0.$ In the same Appendix various choices for
independent parameters of the Higss sector are outlined. Following
ref.\cite
{gunion}, we adopt $\tan \beta $ , $m_{A^0}$ as our indepent parameters,
and $m_{H^0}$, $m_{h^0}$, $\alpha$ can be written as
\begin{eqnarray}
m_{H^0,h^0}^2 &=&\frac 12\left[ m_{A^0}^2+m_Z^2\pm \sqrt{\left(
m_{A^0}^2+m_Z^2\right) ^2-4m_Z^2 m_{A^0}^2 \cos^2 2\beta }\right]
\;\; ,\nonumber \\
\cos 2\alpha &=&-\cos 2\beta \left( \frac{m_{A^0}^2-m_Z^2}{%
m_{H^0}^2-m_{h^0}^2}\right)  \;\; ,\nonumber \\
\sin 2\alpha &=&-\sin 2\beta \left( \frac{m_{H^0}^2+m_{h^0}^2}{%
m_{H^0}^2-m_{h^0}^2}\right) \;\; \text{.}
\end{eqnarray}
\indent The contribution of the 2HDM to the anomalous
chromomagnetic dipole moment of the quark top comes from the exchange
of the 
neutral Higgs $h_o$, $A_o$, and $H_o$ bosons and the would-be Goldstone
boson. The
complete expression is given by
\begin{eqnarray}
\delta \kappa _g^t &=&{\frac{\sqrt{2}G_Fm_t^2}{8\pi ^2}}\!\!\left[ \!({\frac{%
\sin \alpha }{\sin \beta }})^2\!\int_0^1dx{\frac{x-x^3}{x^2-a(m_{H^0}^2)x+1}}%
-({\frac{\cos \alpha }{\sin \beta }})^2\!\int_0^1dx{\frac{x-x^3}{%
x^2-a(m_{h^0}^2)x+1}}\right.  \nonumber \\
&+&\left. \cot ^2\beta \int_0^1dx{\frac{-x^3+2x^2-x}{x^2-a(m_A^2)x+1}}%
+\int_0^1dx{\frac{-x^3+2x^2-x}{x^2-a(m_z^2)x+1}}\right] \; \; .
\end{eqnarray}
In figure 2, we display the contribution in terms of the $A^0$ Higgs mass
for a top mass of $175$ GeV and for some values of $\tan \beta $. There, 
we have already
summed the pure QCD correction.

Our last task is to estimate the anomalous chromomagnetic dipole moment in
the context of the
MSSM \cite{susy}. We only consider
the QCD contribution in the MSSM by
the
supersimetrization of the standard QCD contribution, i.e. 
the exchange of gluinos and squarks. The squark sector presents mixing
between left- and right-handed sectors and this squark
mixing is particularly relevant
in the case of stop because it is proportional to the top quark mass.

If we neglect family mixings, the mixing relation is
given by 
\begin{eqnarray}
\widetilde{t}_a &=&\sum_{b=ij}R_{ab}\widetilde{q}_b  \nonumber \\
R &=&\left( 
\begin{array}{cc}
\cos \theta _t & \sin \theta _t \\ 
-\sin \theta _t & \cos \theta _t
\end{array}
\right) \text{ }
\end{eqnarray}
where the rotation matrix diagonalize the corresponding stop mass matrix \cite{susy} 
\begin{equation}
M_{\widetilde{t}}^2=\left( 
\begin{array}{cc}
m_{\widetilde{t_L}}^2 & \Upsilon m_t \\ 
\Upsilon m_t & m_{\widetilde{t_R}}^2
\end{array}
\right) \;\; , 
\end{equation}
\noindent with the expressions 
\begin{eqnarray}
m_{\widetilde{t_L}}^2 &=&\widetilde{M_Q}+m_Z^2\cos 2\beta \left( \frac 12-%
\frac 23\sin ^2\theta _W\right) +m_t^2 \; ,\\
m_{\widetilde{t_R}}^2 &=&\widetilde{M_U}+m_Z^2\cos 2\beta \left( \frac 23%
\sin ^2\theta _W\right) +m_t^2  \nonumber \; ,\\
\Upsilon &=&\mu \cot \beta +A_t\widetilde{M}  \; \;, \nonumber
\end{eqnarray}
where $\widetilde{M_Q}$, $\widetilde{M_U}$ are the soft supersymmetry
breaking terms, $\mu $ is the coefficient of the mixing Higgs term in the
superpotential and $A_t$ is the trilinear soft SUSY-breaking
parameter \cite
{susy}. In this way, stop $\tilde{t}_1$ and $\tilde{t}_2$ become
the mass
eigenstates, with a
mixing angle 
$\theta _t$ and mass eigenstates given by
\begin{eqnarray}
\tan 2\theta _t &=&\frac{2\Upsilon m_t}{m_{\widetilde{t_L}}^2-m_{\widetilde{%
t_R}}^2}  \nonumber \\
m_{\widetilde{t_{1,2}}} &=&\frac 12\left[ m_{\widetilde{t_L}}^2+m_{%
\widetilde{t_R}}^2\mp \sqrt{\left( m_{\widetilde{t_L}}^2-m_{\widetilde{t_R}%
}^2\right)^2 +4\Upsilon ^2m_t^2}\right] \;\; .
\end{eqnarray}
\indent To evaluate the anomalous chromomagnetic dipole moment in this
framework, we first
consider the exchange of $\tilde{t}_1$. Our result can be
written
as
\begin{eqnarray}
\delta \kappa _g^t &=&{\frac{4\alpha _s}{3\pi }}\int_0^1dx{\ 
%TCIMACRO{
%\dfrac{\cos ^2\theta _t(x^2-3x^3/2)-\sin ^2\theta _t(x^3/2)+\hat{m}_g\sin \theta _t\cos \theta _t(x^2-x)}{x^2-(1-\hat{m}_g^2+\hat{m}_{\tilde{t}_1}^2)x+\hat{m}_{\tilde{t}_1}^2}}
%BeginExpansion
{\displaystyle {\cos ^2\theta _t(x^2-3x^3/2)-\sin ^2\theta _t(x^3/2)+\hat{m}_g\sin \theta _t\cos \theta _t(x^2-x) \over x^2-(1-\hat{m}_g^2+\hat{m}_{\tilde{t}_1}^2)x+\hat{m}_{\tilde{t}_1}^2}}
%EndExpansion
}  \nonumber \\
&&  \nonumber \\
&&+{\frac{3\alpha _s}{2\pi }}\int_0^1dx{\ 
%TCIMACRO{
%\dfrac{\cos ^2\theta _t(x^2-3x^3/2)+\sin ^2\theta _t(x^3/2)+\hat{m}_g\sin \theta _t\cos \theta _t(x-x^2)}{x^2-(1+\hat{m}_g^2-\hat{m}_{\tilde{t}_1}^2)x+\hat{m}_g^2}}
%BeginExpansion
{\displaystyle {\cos ^2\theta _t(x^2-3x^3/2)+\sin ^2\theta _t(x^3/2)-\hat{m}_g\sin \theta _t\cos \theta _t(x^2-x) \over x^2-(1+\hat{m}_g^2-\hat{m}_{\tilde{t}_1}^2)x+\hat{m}_g^2}}
%EndExpansion
}
\end{eqnarray}
where $\hat{m}_i=m_i/m_t$. The expression for the exchange of $\tilde{t}_2$
is obtained by changing $\sin \theta _t$ by $\cos \theta _t$, and
vice-versa.
Figure 3 shows the anomalous
chromomagnetic dipole moment as a function of $\tilde{M}$ with $\mu=-30$ GeV
and $m_g=200$ GeV, for some different values of $\tan \beta $.
In figure 4 we plot the anomalous chromomagnetic dipole moment of the
top quark as a function of $\mu$ with the fixed values
$\tilde{M}=\widetilde{M_Q}=\widetilde{M_U}=500$
GeV, $A_t=0$, $m_g=200$ GeV and $m_t=175$ GeV, and for different values of 
$\tan\beta$ (as in
the other figures). Finally, figure 5 shows $\delta \kappa _g^t$ as
a
function of the gluino mass $m_g$, for the same values of $\tan
\beta$.

In conclusion, we have calculated the chromomagnetic dipole moment of the
top quark in the framework of the SM, the 2HDM with a discrete symmetry
( the so-called model
II), as well as the SUSY-QCD corrections. In all cases, we have added
the pure QCD
correction which is proportional to $\alpha_s$, like QED for the $g-2$ 
factor. In the SM the main contribution comes from the neutral Higgs 
particle.

A naive estimation for the contribution of new physics gives
$\delta\kappa_g^t\approx 0.031$
for $\Lambda=1$ TeV, and $\delta\kappa_g^t\approx 0.25$ for a techniscalar
mass of the order of $0.5$ TeV \cite{atwood}. Using the effective
Lagrangian approach 
in the $b\to s\gamma$ process was found that  $\delta\kappa_g^t\approx
0.066$ \cite{martinez}. In this paper, we have found that the SM
contribution
is of the order
of $-0.004\leq\delta\kappa_g^t\leq
-0.001$. This result is similar to the one  obtained in ref.
\cite{tau} 
for $\sqrt{s}=500$ GeV. The values that we have found for different
scenarios
like 2HDM and SUSY-QCD are of the order of $-0.001\leq\delta\kappa_g^t\leq
-0.01$. In ref. \cite{tg} it was found that $ee \to \bar{t}tg$ processes 
at the Next Linear Collider (NLC) will be sensitive to 
$-0.027\leq\delta\kappa_g^t\leq 0.026$ which will allow to test physics
beyond the SM throught this dipole moment.

We acknowledge helpfull discussions with M.A. Perez and F. Larios. The
present work was complete during a visit to the Physics Department of
CINVESTAV. It was supported by COLCIENCIAS and CONACyT. M. V. thanks
Fundaci\'on Mazda para el Arte y la Ciencia for his fellowship.

\newpage

\begin{center}
Figure Captions
\end{center}

\noindent Figure 1. $\delta\kappa^t_g$ as a function of the Higgs mass. 
Horizontal dashed line is the QCD correction. The solid-line is the SM
Higgs contribution, and dotted is the sum of QCD and Higgs SM contributions.

\vspace{1cm}

\noindent Figure 2. $\delta\kappa_g^t$ as a function of the Higgs mass in
2HDM for three different values of $\tan\beta=1.5, 5$ and $15$ corresponding to solid, dashed and dotted curves, respectively.

\vspace{1cm}

\noindent Figure 3. The same as the fig. 2 for the MSSM contribution as a function of $\tilde{M}$ with $\mu=-30$ GeV, $A_t=0$ and $m_g=200$ GeV.

\vspace{1cm}

\noindent Figure 4. The same as the fig. 3 as a function of $\mu$ instead, with
$\tilde{M}=500$ GeV, $A_t=0$ and $m_g=200$ GeV.

\vspace{1cm}

\noindent Figure 5.  $\delta\kappa^t_g$ as a function of the gluino mass
for 
$\mu=-30$ GeV, $A_t=0$ and $\tilde{M}=500$ GeV. $\tan\beta$ takes the same 
values of the fig. 2.

\newpage

\begin{figure} 
\centerline{\hbox{ 
\psfig{figure=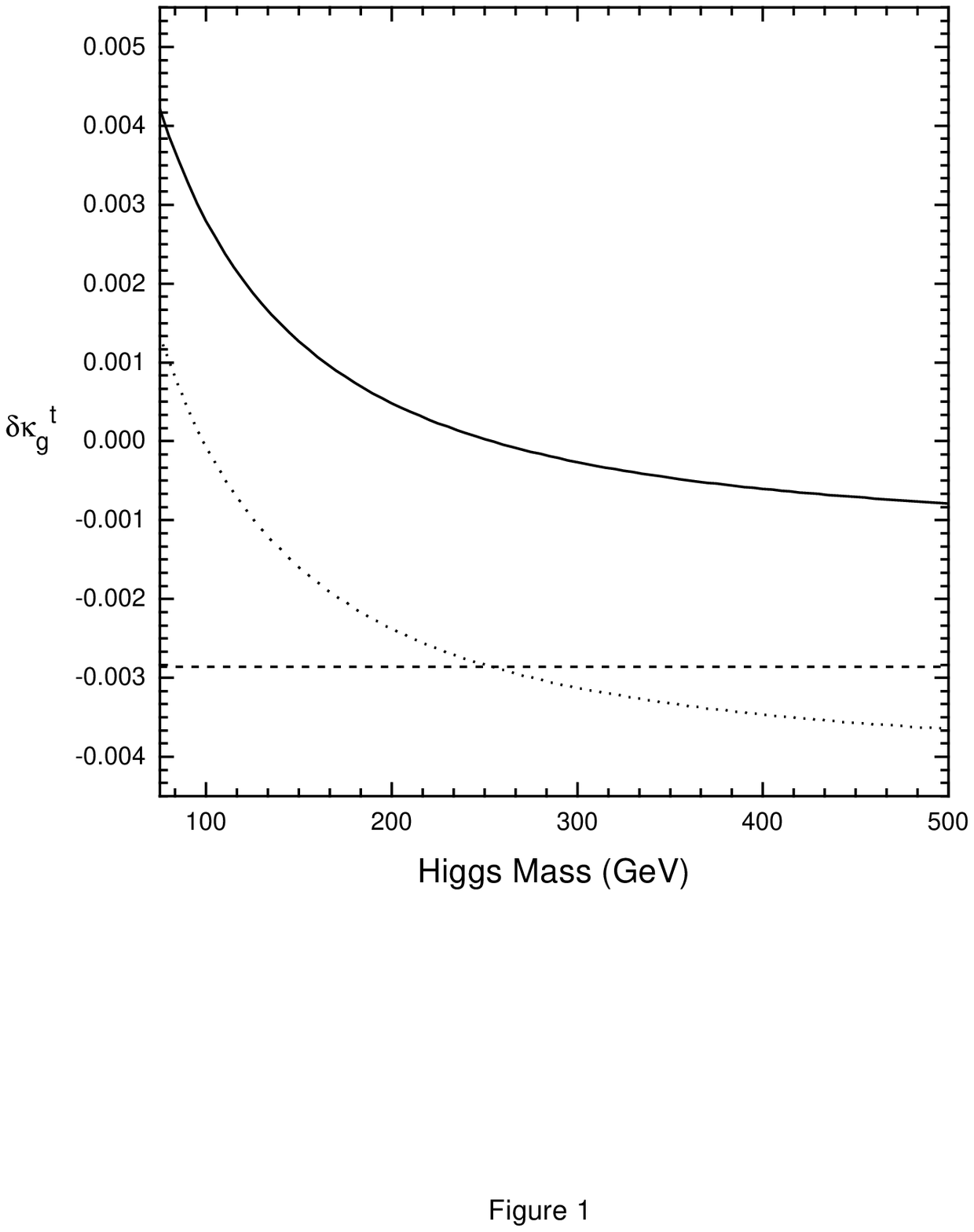,height=1.0in}}}
\end{figure}

\newpage

\begin{figure} 
\centerline{\hbox{ 
\psfig{figure=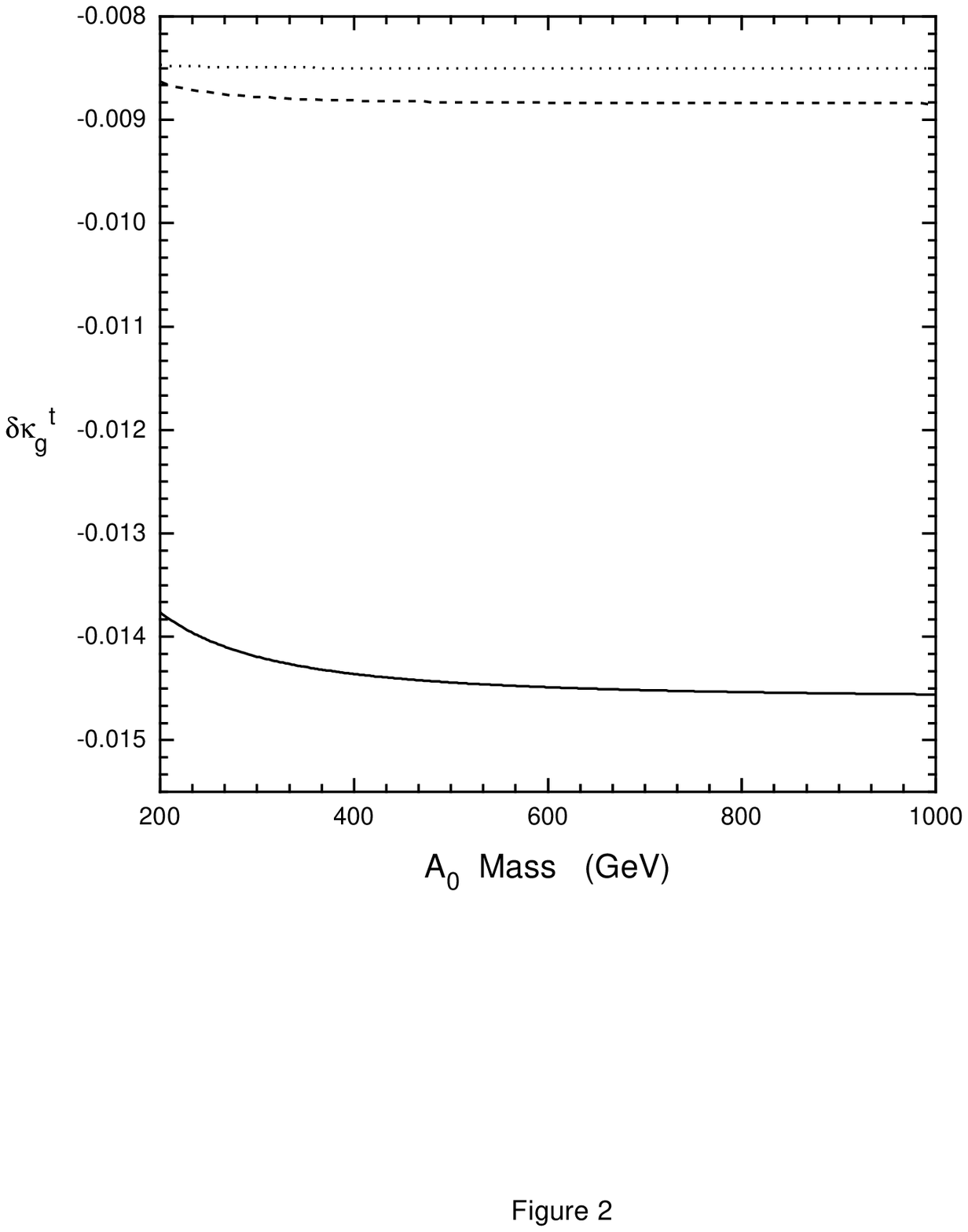,height=1.0in}}}
\end{figure}

\newpage

\begin{figure} 
\centerline{\hbox{ 
\psfig{figure=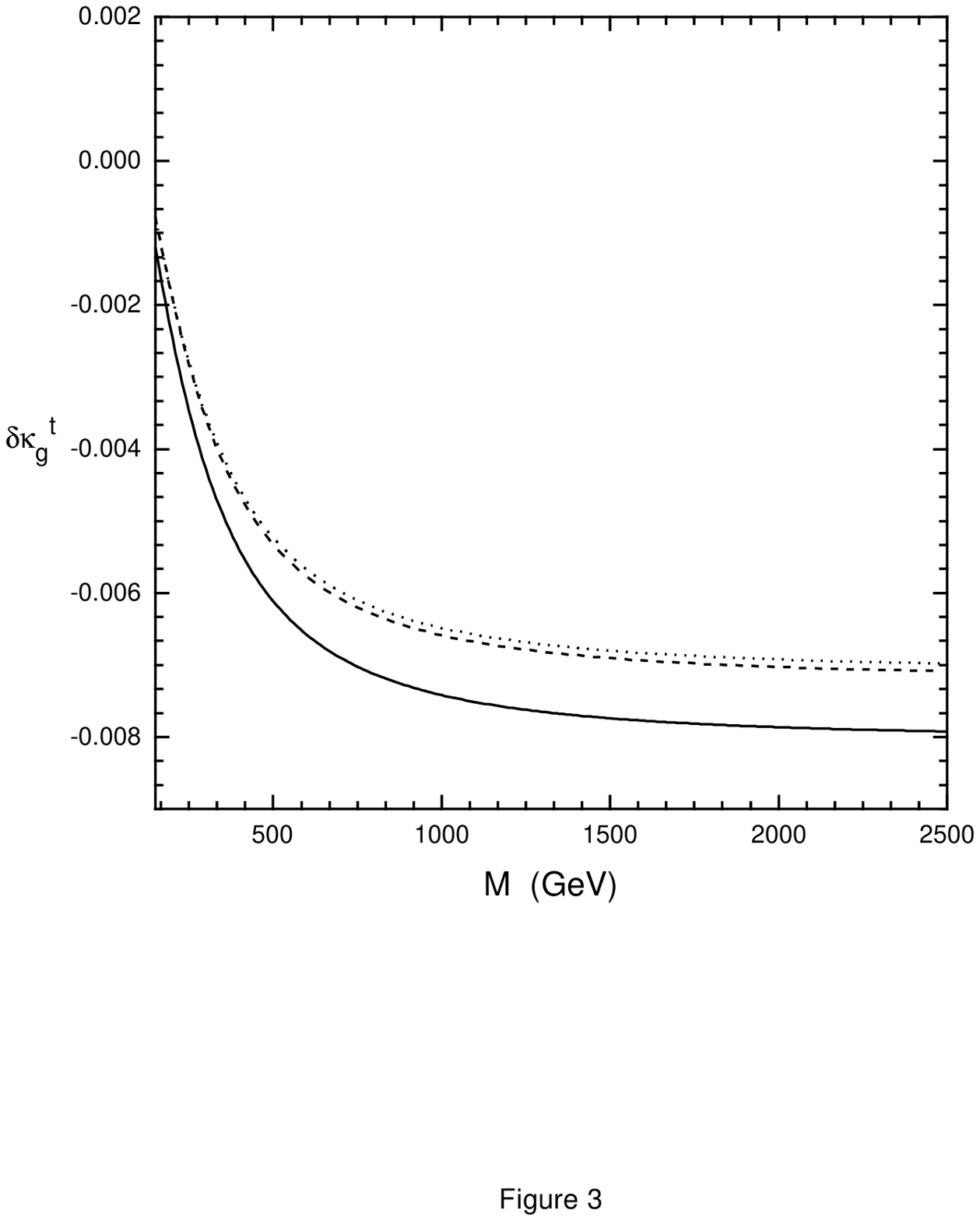,height=1.0in}}}
\end{figure}

\newpage

\begin{figure} 
\centerline{\hbox{ 
\psfig{figure=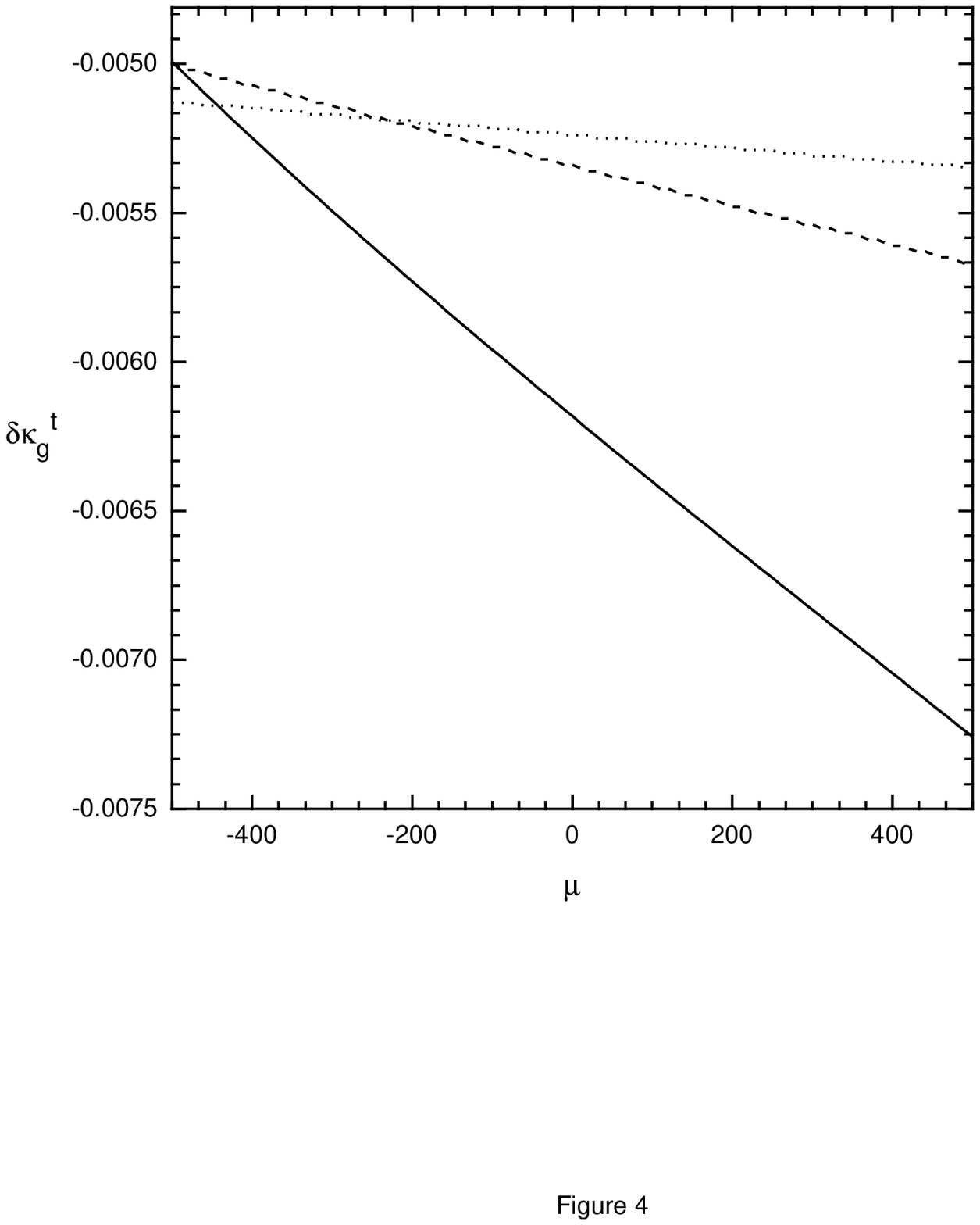,height=1.0in}}}
\end{figure}

\newpage

\begin{figure} 
\centerline{\hbox{ 
\psfig{figure=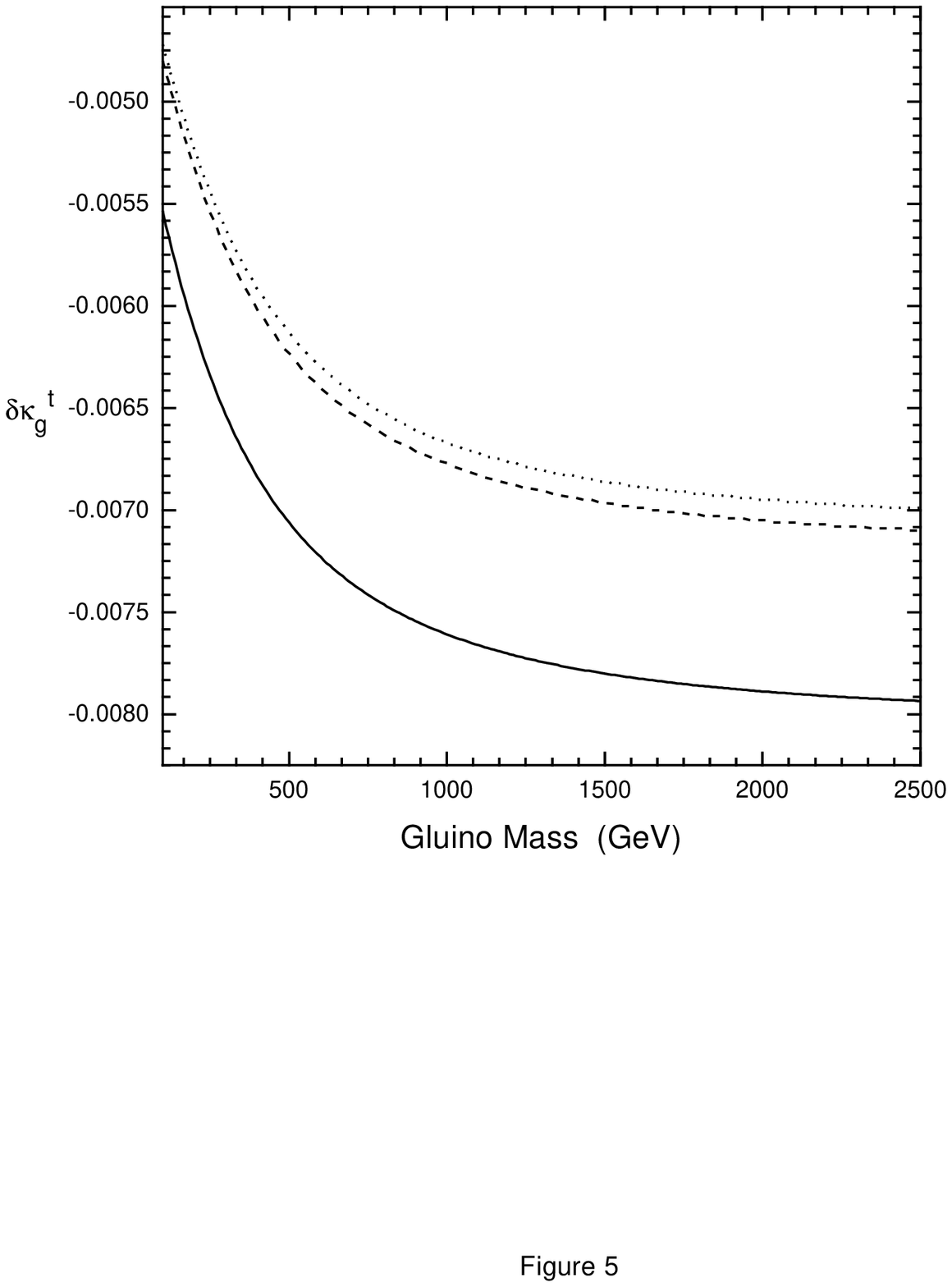,height=1.0in}}}
\end{figure}

\end{document}